\documentclass[lettersize,journal]{IEEEtran}
\usepackage{xcolor} 

\usepackage{amsmath,amsfonts}
\usepackage{algorithmic}
\usepackage{array}
\usepackage[caption=false,font=normalsize,labelfont=sf,textfont=sf]{subfig}
\usepackage{textcomp}
\usepackage{stfloats}
\usepackage{url}
\usepackage{verbatim}
\usepackage{graphicx}
\usepackage{rotating}       
\usepackage{tablefootnote}
\usepackage{pgfplots}
\usepackage{adjustbox} 
\usepackage{makecell} 

\usepackage{array}  
\usepackage{multirow}   

\usepackage{threeparttable}

\usepackage{graphicx}
\usepackage{subfig}
\usepackage[ruled,linesnumbered, noline]{algorithm2e}

\pgfplotsset{compat=1.18}

\def\BibTeX{{\rm B\kern-.05em{\sc i\kern-.025em b}\kern-.08em
    T\kern-.1667em\lower.7ex\hbox{E}\kern-.125emX}}
\usepackage{balance}

\usepackage{caption}
\begin{document}
\title{ECG-LDC: A Hardware-Efficient Low-Dimensional Computing Framework for ECG Arrhythmia Classification}

\author{Anh Tran, Khanh Tran, Cuong Do 
\thanks{Manuscript created June, 2026. This work was supported by the VinUni--Illinois Smart Health Center, VinUniversity, Hanoi 100000, Vietnam, , under Project No.~10000167, titled ``Point of Care and Telehealth Diagnostics for Data-Driven Smart Health Systems.'' \textit{(Corresponding authors: Anh Tran and Cuong Do.)}

Anh Tran is with the School of Engineering and Applied Science, University of Pennsylvania, Philadelphia, PA 19104, USA, and also with the VinUni--Illinois Smart Health Center, VinUniversity, Hanoi 100000, Vietnam (email: anh.thh@vinuni.edu.vn).

Khanh Tran is with the College of Engineering and Computer Science, VinUniversity, Hanoi 100000, Vietnam (e-mail: khanh.tg@vinuni.edu.vn).

Cuong Do is with the College of Engineering and Computer Science, VinUniversity, Hanoi 100000, Vietnam, and also with the VinUni--Illinois Smart Health Center, VinUniversity, Hanoi 100000, Vietnam (e-mail: cuong.dd@vinuni.edu.vn).
}}

\markboth{Preprint, 2026}%
{ECG-LDC: A Hardware-Efficient Framework for ECG Arrhythmia Classification}

\maketitle

\begin{abstract}
Continuous cardiac monitoring in wearable devices demands classifiers that are simultaneously accurate, energy-efficient, and deployable on resource-constrained hardware. While deep neural network approaches have demonstrated high classification accuracy for electrocardiogram (ECG) arrhythmia detection, their substantial parameter counts and reliance on multiply-accumulate-intensive operations make them impractical for low-cost edge platforms. In this work, we propose ECG-LDC, a hardware-software co-design framework that adapts  Low-Dimensional Computing (LDC) for real-time ECG arrhythmia classification. ECG-LDC employs a dual-encoder architecture with dedicated value and feature codebooks to independently encode morphological waveform features and RR-interval temporal features, enabling effective capture of both intra-beat and inter-beat cardiac dynamics. The framework encompasses data preprocessing, model training, and a hardware accelerator architecture prototyped on the Pynq-Z2 platform. Implemented using binary representations and XOR/XNOR-based operations, ECG-LDC achieves $97.18\%$ accuracy with a memory footprint of only $3.86\ \text{kB}$.  ECG-LDC sacrifices approximately $1.8\%$ accuracy versus SOTA TinyML classifiers but achieves $11\text{--}570\times$ reduction in memory usage; among FPGA-based five-class arrhythmia classifiers, it delivers the highest accuracy with up to $2.4\times$ fewer LUTs and zero DSP block utilization, affirming its suitability for real-time arrhythmia detection on resource-constrained wearable platforms. 
\end{abstract}


\section{Introduction}
\IEEEPARstart{C}{ardiovascular} diseases (CVDs) are a group of disorders affecting the heart and blood vessels, encompassing conditions such as coronary artery disease, heart failure, and stroke. CVDs remain the leading cause of mortality worldwide, accounting for an estimated 19.2 million deaths annually \cite{doi:10.1016/j.jacc.2025.08.015}. A critical component of CVD treatment and clinical management is the electrocardiogram (ECG), a non-invasive tool that records the heart’s electrical activity using electrodes placed on the body surface. The ECG captures subtle variations in cardiac electrical signals that reflect different heart conditions. It provides rich, insightful information for accurate detection of arrhythmias, i.e., abnormal heartbeats, facilitating early diagnosis of potential cardiac diseases.

In recent years, deep neural networks have emerged as effective solutions for ECG signal analysis. Various approaches leveraging residual connections, convolutional architectures, and attention mechanisms have been proposed, with works such as Allam et al.~\cite{ALLAM20201446} and Islam et al.~\cite{ISLAM2024106211} demonstrating classification accuracies exceeding 99\% on the MIT-BIH Arrhythmia Database~\cite{932724}. However, such approaches come with substantial computational cost. The aforementioned models contain 13.6 million and 1.18 million parameters, respectively, demanding significant memory and processing resources. Consequently, such models are typically deployed on high-end GPUs, dedicated medical workstations, or cloud-based servers, making them impractical for resource-constrained edge platforms such as low-cost wearable devices or implantable cardiac monitors. 

With the rising demand for wearable healthcare monitoring, deploying accurate arrhythmia detection algorithms at the edge has become a critical research challenge. Dedicated hardware accelerators, such as FPGAs and ASICs, offer a compelling solution by customizing the underlying logic to efficiently execute neural network inference, enabling ultra-low power operation and compact form factors well-suited for wearable platforms. However, deploying multi-class ECG classification on such hardware imposes a strict trade-off: high-precision architectures demand substantial multiply-accumulate (MAC) units and on-chip memory that exceed the capacity of low-cost edge devices, while resource-minimized designs often suffer from increased latency or degraded accuracy in detecting complex arrhythmias.

Low-Dimensional Computing (LDC) \cite{duan2022braininspiredlowdimensionalcomputingclassifier} is an emerging lightweight machine learning paradigm designed for deployment on resource-constrained, low-power devices. It represents input features as compact binary vectors and performs classification through simple vector operations and XOR/XNOR computations. Unlike conventional deep learning architectures, which rely on convolutional or attention mechanisms and require extensive MAC operations, LDC employs logic-based computation, thereby inherently minimizing computational complexity, memory overhead, and hardware resource requirements. This eliminates dependence on DSP-intensive arithmetic units, enabling efficient inference on edge and embedded platforms.

Motivated by these properties, this work proposes ECG-LDC, a hardware-software co-design framework that adapts LDC for real-time, multi-class ECG arrhythmia classification under the stringent resource and latency constraints of embedded medical platforms. The primary contributions of this work are summarized as follows:
\begin{enumerate}
    \item We propose a dual-encoder LDC architecture that employs distinct value and feature hypervectors to independently encode morphological and RR-interval features of ECG signals.
    \item We present a hardware architecture for ECG-LDC, prototyped on the Pynq-Z2 FPGA platform, demonstrating its feasibility for low-power, real-time edge deployment.
    \item We comprehensively evaluate ECG-LDC on both classification performance and hardware efficiency, demonstrating its competitiveness against state-of-the-art TinyML ECG classifiers.
\end{enumerate}

The remainder of this paper is organized as follows: Section II provides the necessary background on ECG signals and the mathematical foundations of Low-Dimensional Computing (LDC). Section III details the proposed methodology, including the preprocessing pipeline, the dual-encoder ECG-LDC architecture, its hardware realization, and a comprehensive complexity analysis. Section IV presents the experimental framework, covering the setup, evaluation metrics, and an in-depth discussion of both classification performance and hardware efficiency. Finally, Section V concludes the paper by summarizing the key findings, limitations, and future directions.

\section{Background}
\subsection{Electrocardiogram Data}
\begin{figure}[!t]
    \centering
    \includegraphics[width=\columnwidth]{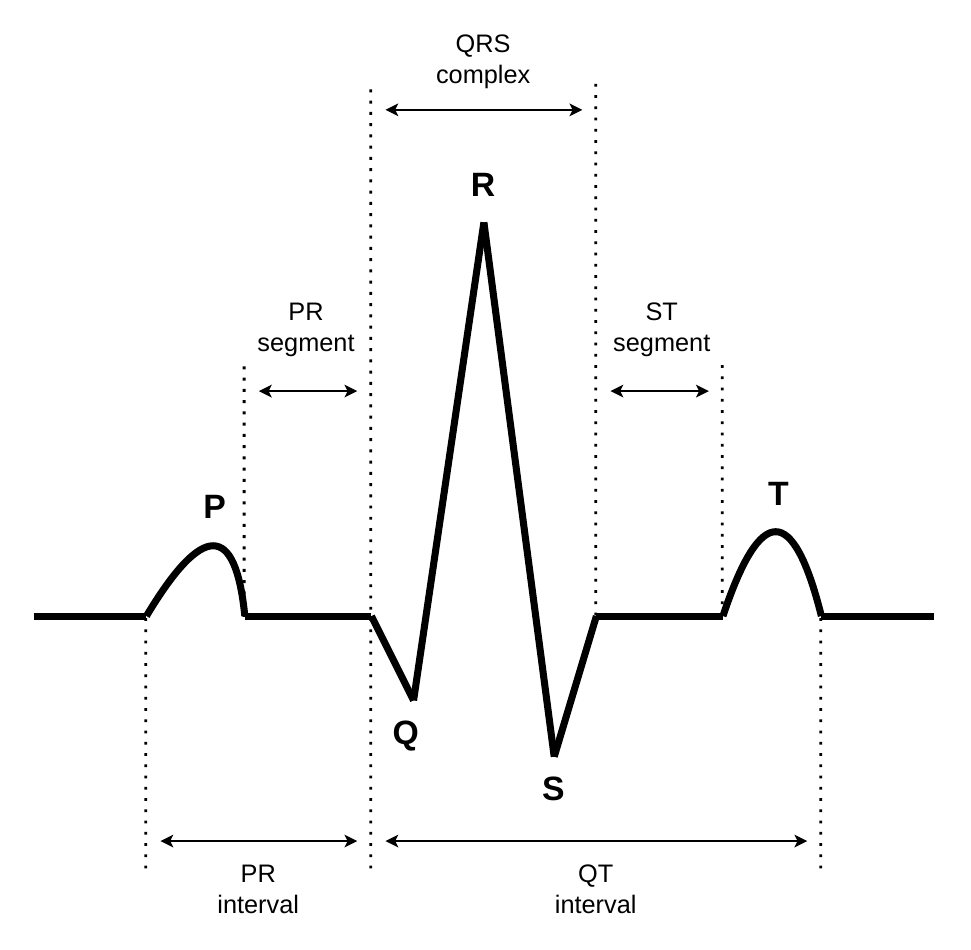}
    \caption{Structure of an ECG beat.}
    \label{fig:ecg}
\end{figure}

An electrocardiogram (ECG) is a biomedical test that records the electrical signal of the heart over time. It is acquired by placing electrodes on the skin surface at standardized positions such as limbs and chest. These electrodes detect the small voltage variations generated by the depolarization and repolarization of cardiac muscle cells during each heartbeat cycle. The resulting signal is amplified and recorded as a continuous waveform, typically sampled at rates ranging from $128\ \text{Hz}$ to $500\ \text{Hz}$ depending on the clinical application. Due to its simplicity, low cost, and rich diagnostic information, the ECG remains the primary tool for detecting and monitoring cardiac arrhythmias in clinical practice.

The ECG signal is characterized by a recurring pattern known as the QRS complex. As illustrated in Figure~\ref{fig:ecg}, the QRS complex consists of three consecutive waves: the Q wave (initial downward deflection), the R wave (dominant upward peak), and the S wave (subsequent downward deflection). The QRS complex typically spans $74$--$114$ ms in duration (average 95 ms in healthy adult males) \cite{doi:10.1161/CIRCULATIONAHA.108.191095}. Its morphological structure serves as the main reference point for heartbeat segmentation and classification.

Another important feature extracted from the ECG is the RR-interval, defined as the time between two consecutive R-wave peaks. As the R peak is the highest-amplitude point in the QRS complex and one of the most reliably detectable, the RR-interval directly measures instantaneous heart rate and its variability. Under normal conditions, RR-intervals typically range from $0.6$--$1$ seconds, corresponding to a heart rate of $60$--$100$ beats per minute \cite{Hafeez2023}. RR-intervals convey information about the global temporal relationship between heartbeats; deviations from normal patterns often indicate various cardiac abnormalities.

\subsection{Low-Dimensional Computing}
\label{sec:ldc}
Consider an input vector $\mathbf{x} = [x_0,\ x_1,\ \dots,\ x_{N-1}]$ with $N$ discrete-valued features, where $x_i \in \{0,1,\dots,M-1\}$. LDC represents $\mathbf{x}$ using two bipolar codebooks: a value codebook $\mathbf{V} \in \{-1,+1\}^{M \times d_v}$ and a feature codebook $\mathbf{F} \in \{-1,+1\}^{N \times d_f}$, where each row corresponds to a value level and feature position, respectively. For each feature $x_i$, the associated value hypervector $\mathbf{V}[x_i]$ is bound to the corresponding feature hypervector $\mathbf{F}[i]$ via element-wise multiplication. The resulting bound hypervectors are accumulated and bipolarized using the sign function:
\begin{align}
    \mathbf{e} = \operatorname{sign}\left(\sum_{i=0}^{N-1} \mathbf{V}[x_i] \odot \mathbf{F}[i]\right) 
\end{align}
The output $\mathbf{e} \in \{-1,+1\}^{d_f}$ is the bipolar embedding representing the full input $x$. Note that the dimensions $d_v$ and $d_f$ are not necessarily equal; in practice, $d_v$ is usually smaller, and the value embedding $\mathbf{V}[x_i]$ can be duplicated to match the dimension of $\mathbf{F}[i]$.

To classify the embedding $\mathbf{e}$, the LDC model employs a class codebook $\mathbf{C} \in \{-1,+1\}^{K \times d_f}$, where each row corresponds to a class hypervector. The predicted class is determined as the one whose hypervector is closest to $\mathbf{e}$ according to the Hamming distance:
\begin{align}
    \hat{y} = \underset{k}{\operatorname{arg\,min}}\ \mathrm{Hamm}(\mathbf{e}, \mathbf{C}[k]),
\end{align}
where the Hamming distance is defined as
\begin{align}
    \mathrm{Hamm}(\mathbf{e}_1, \mathbf{e}_2)
    = \frac{1}{d_f} \sum_{j=1}^{d_f}
    \mathbb{I}\!\left[(\mathbf{e}_1)_j \neq (\mathbf{e}_2)_j\right].
\end{align}

A key strength of LDC is its ability to learn the codebooks $\mathbf{V}$, $\mathbf{F}$, and $\mathbf{C}$ through gradient-based optimization. To enable end-to-end training, the discrete codebooks are parameterized as:
\begin{align}
    \mathbf{V}[x_i] &= \operatorname{sign}_{\mathrm{STE}} \!\left( f_{\tilde{\theta}}(x_i) \right), \\
    \mathbf{F} &= \operatorname{sign}_{\mathrm{STE}} \!\left( \tilde{\mathbf{F}} \right), \\
    \mathbf{C} &= \operatorname{sign}_{\mathrm{STE}} \!\left(\tilde{\mathbf{C}} \right),
\end{align}
where $f_{\tilde{\theta}}(\cdot)$ denotes a trainable nonlinear mapping, while $\tilde{\mathbf{F}}$ and $\tilde{\mathbf{C}}$ represent the underlying floating-point tensors. Simulated bipolar quantization is performed using the straight-through estimator (STE),
\begin{align}
    \operatorname{sign}_{\mathrm{STE}}(\mathbf{z}) = \operatorname{sign}(\mathbf{z}) - \operatorname{sg}(\mathbf{z}) + \mathbf{z},
\end{align}
where $\operatorname{sg}(\cdot)$ denotes the stop-gradient operator. 

On hardware, LDC models are highly efficient, as bipolar representations can be directly converted to binary form, enabling compact, memory-efficient implementations. In addition, bipolar arithmetic can be realized using low-cost XOR/XNOR operations, while the element-wise encoding stage naturally exposes parallelism for low-latency FPGA inference. These properties make LDC well-suited for wearable applications operating under tight resource and power constraints.

\section{Methodology}

\subsection{Preprocessing}
\begin{figure*}[!t]
    \centering
    \includegraphics[width=0.85\textwidth]{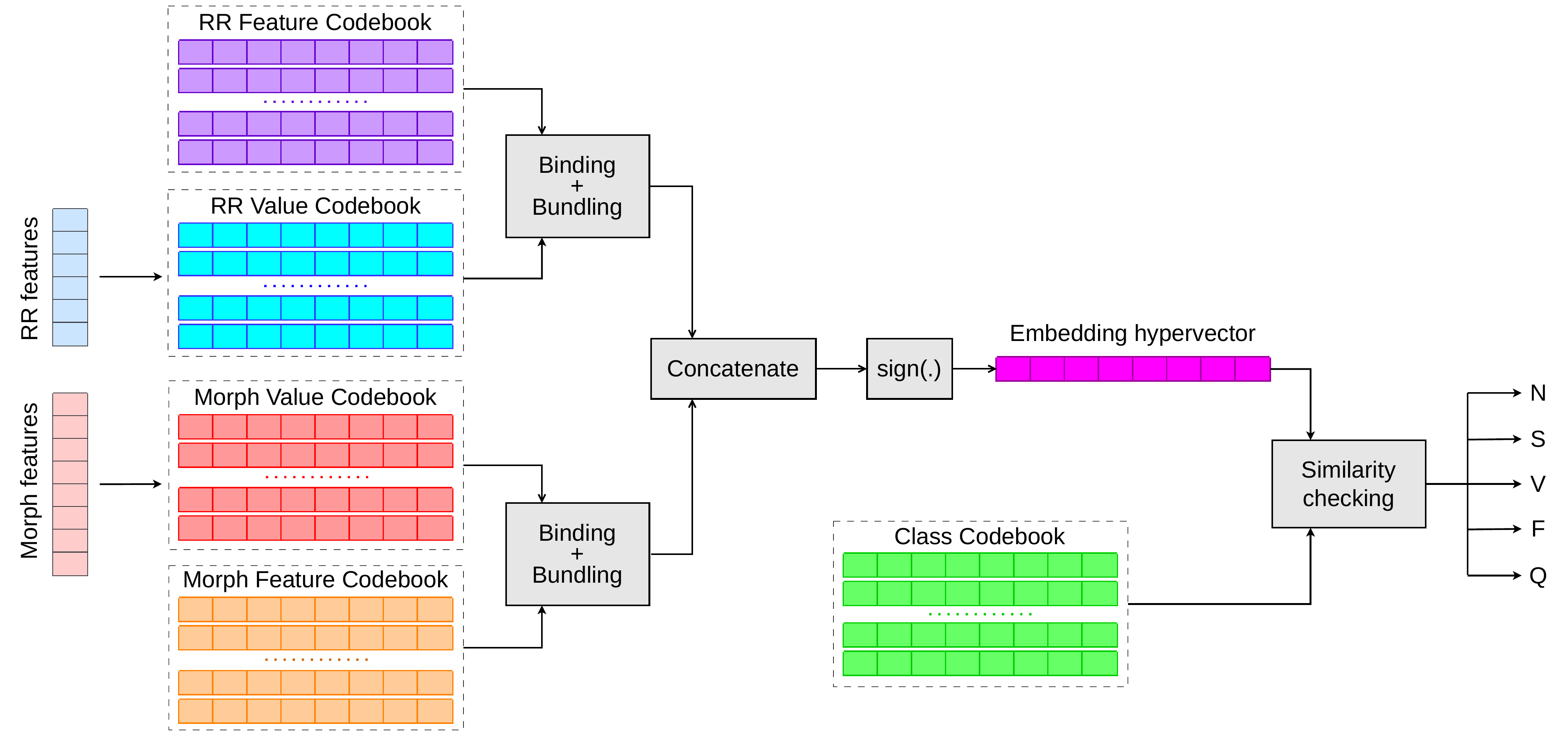}
    \caption{ECG-LDC architecture overview.}
    \label{fig:ecg-ldc}
\end{figure*}

For data preprocessing, we adopt the Pan--Tompkins algorithm \cite{4122029, ZISHAN2024122560}, which is widely used to amplify QRS complexes in ECG signals. The algorithm typically involves a four-step procedure: (1) band-pass filtering, (2) derivative filtering, (3) squaring, and (4) moving window integration. This process reduces noise in the raw ECG signal, enhances the QRS complex, and facilitates the detection of R-peaks for each heartbeat. To reduce input dimensionality, the signal is resampled to 250 Hz, and a window of $N_{m}=61$ samples centered around each R-peak is extracted from the preprocessed signal. The resulting segments capture the morphological features of individual heartbeats. To further incorporate temporal information, each beat $i$ is augmented with an RR-interval feature vector $[RR_i,\ RR_{i+1},\ \overline{RR}_i,\ RR_{wSDNN},\ RR_{index}]$, where the first three entries represent the RR interval between the current beat and its preceding beat, the RR interval between the current beat and its succeeding beat, and the local mean RR interval computed over 11 consecutive intervals from $RR_{i-9}$ to $RR_{i+1}$, respectively. The remaining two features are defined as
\begin{align}
    RR_{wSDNN} = \sqrt{\dfrac{1}{10}\sum_{j=-9}^{1} w_j (RR_{i+j} - \overline{RR}_i)^2}
\end{align}
where $w_j = 10$ if $j = 0$ and $w_j = 1$ otherwise; and
\begin{align}
    RR_{index} = 2\ \dfrac{RR_i - RR_{i-1}}{RR_i + RR_{i-1}}
\end{align}

As LDC models require discrete-valued input vectors, linear quantization is applied to map floating-point signals to $b$-bit integers in the range $[0, 2^{b}-1]$:
\begin{align}
    x_{\text{quant}} = \mathrm{round} \left( \dfrac{x - x_{\min}}{x_{\max} - x_{\min}} \cdot (2^{b} - 1) \right),
\end{align}
with $b = 8$ in this work. The quantization parameters $x_{\min}$ and $x_{\max}$ are determined from the training dataset and kept fixed afterward.

\subsection{ECG-LDC Architecture for Arrhythmia Classification}
The architecture of the ECG-LDC model is illustrated in Fig.~\ref{fig:ecg-ldc}. 
Since morphological patterns and RR intervals convey complementary information, with the former describing local waveform shapes and the latter reflecting temporal rhythm characteristics, they are processed separately. Using dedicated value and feature codebooks for each modality allows ECG-LDC to learn feature-specific embedding distributions while reducing representational interference that may arise from a shared encoding space. For morphological information, the morph value and feature codebooks have dimensionalities $2^{b} \times d_{v,m}$ and $N_m \times d_{f,m}$, respectively, where $N_m$ denotes the number of morphological features. Analogously, for RR-interval information, the corresponding codebooks have dimensionalities $2^{b} \times d_{v,r}$ and $N_r \times d_{f,r}$, respectively, where $N_r$ denotes the number of RR features. Each feature group is independently encoded using the binding-and-bundling scheme. The resulting hypervectors are then bipolarized and concatenated to form the final joint representation with dimensionality $d_{f,m} + d_{f,r}$. Accordingly, the class codebook is extended to match this size. 

The ECG-LDC model is trained following the procedure described in Section~\ref{sec:ldc}. The nonlinear mapping used to parameterize the value encoding consists of stacked fully connected layers with SiLU activations and dropout. Moreover, as suggested in \cite{duan2025vectoroptimizationlowdimensionalvector}, Batch Normalization (BN) can be incorporated to stabilize training. In particular, a BN layer is applied prior to the bipolarization step during bundling. Following this, the encoding process during training can be written as
\begin{align}
    \mathbf{h}_g &= \sum_i \mathbf{V}_g[x_i] \odot \mathbf{F}_g[i], \\
    \mathbf{e}_g &= \operatorname{sign}_{\mathrm{STE}} \!\left( \operatorname{BN}(\mathbf{h}_g) \right),\label{eq:sign-bn}
\end{align}
where $g \in \{m, r\}$.

\subsection{Hardware Realization of ECG-LDC}
\begin{figure*}[!t]
    \centering
    \includegraphics[width=0.9\textwidth]{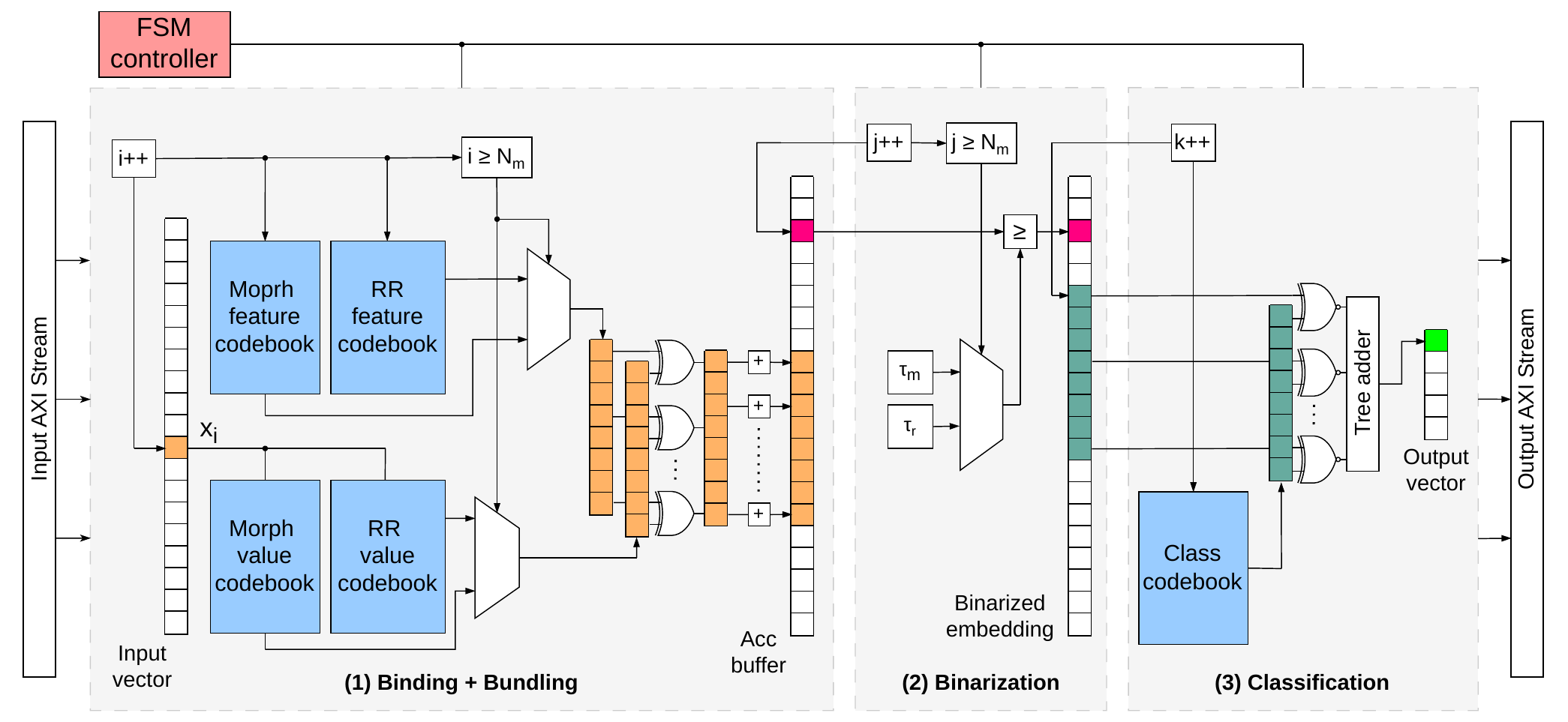}
    \caption{Hardware architecture of the ECG-LDC accelerator.}
    \label{fig:ecg-ldc-hw}
\end{figure*}

In this section, we describe the hardware architecture of ECG-LDC. Unless otherwise specified, bipolar representations are mapped to binary ($\{-1, +1\} \mapsto \{0, 1\}$), with multiplication implemented as XOR and similarity check performed via XNOR-popcount. 
As shown in Fig.~\ref{fig:ecg-ldc-hw}, the accelerator consists of three main modules: (1) Binding \& Bundling, (2) Binarization, and (3) Classification, orchestrated sequentially by a finite state machine (FSM) controller. An input buffer stores incoming ECG samples, and inference begins once a complete input vector has been collected; a similar buffer holds the output class scores.
The codebooks are kept in on-chip block RAM (BRAM). While inferring codebooks as lookup tables (LUTs) is possible and would reduce access latency, such an approach may significantly increase logic utilization as the codebook size grows; BRAM-based storage therefore offers a more scalable and resource-efficient trade-off at larger scales. The accelerator communicates with external systems via an AXI-Stream interface, reflecting the sample-by-sample nature of ECG acquisition. Input data are fed from upstream components, such as a CPU or sensors, into the accelerator for processing.

(1) Binding \& Bundling: The pseudo-code for this stage is presented in Algorithm~\ref{alg:binding_bundling}. In the hardware implementation, multiplexers (MUXes) are employed to select the appropriate memory partition, corresponding to either the morphological or RR codebooks, based on an index counter that sequentially addresses the hypervectors in BRAM. The unified loop structure reflects the shared-datapath hardware implementation. Although the inner loop iterates up to $\max(d_{f,m}, d_{f,r})$, the conditional checks ensure that computations are performed only over the valid dimensions of each feature group; the effective computational complexity is discussed later in this section.
\begin{algorithm}
\caption{Binding and Bundling}
\label{alg:binding_bundling}
\KwIn{$\mathbf{x},\, \mathbf{V}_m,\, \mathbf{F}_m,\, \mathbf{V}_r,\, \mathbf{F}_r$}
\For{$i \leftarrow 0$ \KwTo $N_m + N_r - 1$}{
    \For{$j \leftarrow 0$ \KwTo $\max(d_{f,m},\, d_{f,r}) - 1$}{
        \If{$i < N_m$ \textbf{and} $j < d_{f,m}$}{
            $a \leftarrow \mathbf{V}_m[\mathbf{x}[i]][j \bmod d_{v,m}] \oplus \mathbf{F}_m[i][j]$\;
            $\text{acc}[j] \mathrel{+}= \begin{cases} -1 & \text{if } a = 1 \\ +1 & \text{otherwise} \end{cases}$\;
        }
        \If{$i \geq N_m$ \textbf{and} $j < d_{f,r}$}{
            $a \leftarrow \mathbf{V}_r[\mathbf{x}[i]][j \bmod d_{v,r}] \oplus \mathbf{F}_r[i - N_m][j]$\;
            $\text{acc}[j + d_{f,m}] \mathrel{+}= \begin{cases} -1 & \text{if } a = 1 \\ +1 & \text{otherwise} \end{cases}$\;
        }
    }
}
\end{algorithm}

To facilitate parallel processing, data in BRAM is packed into words of $P$ bits wide. This allows the inner loop (line 2 of Algorithm~\ref{alg:binding_bundling}) to be unrolled by a factor of $P$, such that the XOR and accumulation operations are executed simultaneously across $P$ dimensions. As a result, the cycle latency is reduced, at the cost of additional parallel adders and XOR gates to handle the partial sums. 

(2) Binarization: This stage converts each element of the accumulation vector to a binary value. Since all BN parameters are fixed after training, the $\operatorname{sign}(\operatorname{BN}(\cdot))$ operation in Equation~\eqref{eq:sign-bn} can be simplified. The BN function is defined as:
\begin{align}
    \operatorname{BN}(\mathbf{z}) = \frac{\mathbf{z} - \mu}{\sqrt{\sigma^2 + \epsilon}} \cdot w_{BN} + b_{BN}
\end{align}
where $\mu$ and $\sigma^2$ are the running mean and variance. Taking the sign of $\operatorname{BN}(\mathbf{z})$ and rearranging, the binarization reduces to a single integer comparison against a precomputed threshold:
\begin{align}
    \tau = \mu - \frac{b_{BN} \cdot \sqrt{\sigma^2 + \epsilon}}{w_{BN}}
\end{align}
The threshold, either $\tau_m$ or $\tau_r$, is selected based on whether the current accumulator element belongs to the morphological or RR partition.

(3) Classification: This stage computes the similarity between the binary embedding and each class hypervector by counting matching bits. Codebook entries are packed into $P$-bit words to enable parallel memory reads, with bit-wise XNOR followed by a tree-structured adder for popcount. The class with the highest score is selected as the predicted label.

\subsection{Complexity Analysis}

Let $N=N_m+N_r$ denote the total number of input features,  $d=d_{f,m}+d_{f,r}$ the final embedding dimension, and $N_c$ the number of classes. The size of ECG-LDC is primarily determined by the value, feature, and class codebooks. Thus, the model size complexity is

\begin{align}
    M_{\text{model}} =
    \mathcal{O}\Big(
    &2^b(d_{v,m}+d_{v,r}) + N_m d_{f,m} \nonumber \\
    &+ N_r d_{f,r} + N_c(d_{f,m}+d_{f,r})
    \Big).
\end{align}

The runtime memory overhead, including the input buffer, accumulator, binary embedding buffer, and output score buffer, is estimated as

\begin{align}
    M_{\text{runtime}} = \mathcal{O}\Big( N b_{\text{in}} + d b_{\text{acc}}+ d + N_c b_{\text{out}}\Big).
\end{align}

The computational cost of inference is dominated by two stages: binding-and-bundling and classification. Their complexities are given by

\begin{align}
    C_{\text{bind}+\text{bundling}}
    &=
    \mathcal{O}\left( N_m d_{f,m} + N_r d_{f,r}
    \right), \label{eq:model_size} \\
    C_{\text{classification}}
    &= \mathcal{O}\left( N_c(d_{f,m}+d_{f,r})
    \right). \label{eq:mem_overhead}
\end{align}

Therefore, the total inference complexity is
\begin{align}
    C_{\text{inference}} =
    \mathcal{O}\Big(
    &N_m d_{f,m}
    + N_r d_{f,r} \nonumber \\
    &+ N_c(d_{f,m}+d_{f,r})
    \Big).
\end{align}

With a packed word width of $P$, the binding, bundling, and classification stages are parallelized across $P$ bits. Including the overhead of input loading, binarization, and output delivery, the effective cycle latency becomes

\begin{align}
    T_{\text{cycle}} = \mathcal{O}\Big(
    &N
    + \frac{N_m d_{f,m} + N_r d_{f,r}}{P} \nonumber \\
    &+ d
    + \frac{N_c d}{P}
    + N_c
    \Big).
\end{align}

The computation cost of ECG-LDC scales linearly with the number of input features, embedding dimension, and number of classes. 

\section{Experiments}
\label{sec:experiments}
\subsection{Dataset}
We evaluate our method on the MIT-BIH Arrhythmia Database, a widely used benchmark for ECG arrhythmia classification. The database contains 48 half-hour two-channel ECG recordings from 47 subjects with a sampling rate of 360 Hz with 11-bit resolution over a 10 mV range. Annotations were independently reviewed by two or more cardiologists, resulting in approximately 110000 labeled beat instances. 

According to the Association for the Advancement of Medical Instrumentation (AAMI), there are 15 arrhythmia types, which are mapped into five heartbeat classes: Normal beats (N), Supraventricular ectopic beats (S), Ventricular ectopic beats (V), Fusion beats (F), and Unknown beats (Q). In line with a large body of arrhythmia classification literature, this work adopts this standardized five-class heartbeat grouping. Only the first channel of the signal is fed to the neural network. The dataset is randomly split into 70\% for training and 30\% for testing. All reported results are evaluated on the held-out test set.

\subsection{Experimental Setup}
\label{sec:setup}
The design parameters of ECG-LDC used in our experiments are summarized in Table~\ref{tab:params}.
\begin{table}[!t]
    \caption{ECG-LDC Design Parameters}
    \label{tab:params}
    \centering
    \renewcommand{\arraystretch}{1.2}
    \begin{tabular}{clc}
        \hline
        \textbf{Parameter} & \textbf{Description} & \textbf{Value} \\
        \hline
        $N_m$             & Number of morphological features       & 61  \\
        $N_r$             & Number of RR-interval features         & 5 \\
        $N_c$             & Number of classes                      & 5 \\
        $d_{v,m}$         & Morphological value hypervector size   & 16 \\
        $d_{v,r}$         & RR-interval value hypervector size     & 16  \\
        $d_{f,m}$         & Morphological feature hypervector size & 256 \\
        $d_{f,r}$         & RR-interval feature hypervector size   & 256 \\
        $b_{\text{in}}$   & Input bit-width                        & 8 \\
        $b_{\text{out}}$  & Output bit-width                       & 16 \\
        $P$               & Unroll factor (packed word width)      & 64 \\
        \hline
    \end{tabular}
\end{table}
The model is implemented using the PyTorch framework and trained on an NVIDIA A100 GPU. The Adam optimizer is used with an initial learning rate of $1\times10^{-3}$ and a weight decay of $1\times10^{-4}$. The learning rate is halved at every epoch using a step scheduler. Cross-entropy loss is adopted as the objective function for multiclass classification. The model is trained for $100$ epochs with a batch size of $128$, with each  run taking approximately $15$--$20$ minutes. 

The hardware realization of ECG-LDC is developed in SystemVerilog and implemented using the Vivado 2025.2 toolchain. The model is deployed on the PYNQ-Z2 board, where the ECG-LDC accelerator is mapped to the 28nm Programmable Logic (PL) fabric. Data transfer between the accelerator and the dual ARM Cortex-A9 Processing System (PS) is managed via AXI Direct Memory Access (DMA), which streams ECG input data and retrieves the output class scores through the AXI-Stream interface. The DMA engine is controlled by the PS through an AXI-Lite interface.

\subsection{Evaluation Metrics}
To demonstrate the classification performance of ECG-LDC, we employ five metrics: accuracy (Acc) -- the overall correctness of the predictions; precision (Prec) -- the correctness of positive predictions; sensitivity (Sens) -- the proportion of positive instances correctly identified; specificity (Spec) -- the proportion of negative instances correctly identified; and F1-score -- the harmonic mean of precision and sensitivity. These metrics are formally defined as:
\begin{align}
    \mathrm{Acc} &= \frac{TP + TN}{TP + TN + FP + FN} \\
    \mathrm{Prec} &= \frac{TP}{TP + FP} \\
    \mathrm{Sens} &= \frac{TP}{TP + FN} \\
    \mathrm{Spec} &= \frac{TN}{TN + FP} \\
    \mathrm{F1} &= \dfrac{2 \cdot \mathrm{Prec} \cdot \mathrm{Sens}}{\mathrm{Prec} + \mathrm{Sens}}
\end{align}
where $TP$, $TN$, $FP$, and $FN$ denote true positives, true negatives, false positives, and false negatives, respectively.

In addition, we report the receiver operating characteristic (ROC) curve, which illustrates the trade-off between the true positive rate (TPR) and the false positive rate (FPR), and the precision-recall (PR) curve, which captures the trade-off between precision and sensitivity. These curves are summarized by the area under the ROC curve (AUROC) and the average precision (AP), respectively.

The model size and memory usage are estimated based on the model parameters and runtime buffer overheads described in Equations~\eqref{eq:model_size} and~\eqref{eq:mem_overhead}. These two factors determine whether the model can fit within the limited RAM and storage of edge devices.
Inference latency and throughput, which are measured on the target device, are evaluated to ensure that the classifier operates in real time, specifically sustaining a processing rate sufficient to keep pace with the incoming ECG sample feed rate. Power consumption, which is obtained from the post-implementation report, is also considered, as it directly impacts battery life and thermal constraints in wearable form factors.
Finally, since FPGA is the primary target platform of this work, resource utilization is reported across four categories: LUTs, Flip-Flops (FFs), Block RAM (BRAM), and Digital Signal Processing (DSP) units. Higher resource consumption correlates with increased area, delay, and power dissipation, and a practical wearable ECG solution must fit within the resource budget of the target device.

\subsection{Classification Performance}

\begin{table}[h]
\centering
\caption{Confusion matrix}
\renewcommand{\arraystretch}{1.5}
\setlength{\tabcolsep}{6pt}
\begin{tabular}{cc|c|c|c|c|c|c}
\multicolumn{1}{c}{} &
\multicolumn{6}{c}{\textbf{Predicted Label}} \\
\multirow{6}{*}{\rotatebox{90}{\textbf{True Label}}}
&   & \textbf{N} & \textbf{S} & \textbf{V} & \textbf{F} & \textbf{Q} \\
\cline{2-7}
& \textbf{N} & 26969 & 84  & 90   & 8   & 39   \\
\cline{2-7}
& \textbf{S} & 255   & 540 & 39   & 0   & 0    \\
\cline{2-7}
& \textbf{V} & 167   & 12  & 1957 & 18  & 17   \\
\cline{2-7}
& \textbf{F} & 69    & 0   & 5    & 167 & 0    \\
\cline{2-7}
& \textbf{Q} & 120   & 1   & 2    & 0   & 2290 \\
\cline{2-7}
\label{tab:confusion-matrix}
\end{tabular}
\end{table}
Table~\ref{tab:confusion-matrix} presents the confusion matrix of ECG-LDC on the test set, illustrating the prediction distribution across all five arrhythmia classes. Based on the classification results, the proposed model achieves an overall accuracy of $97.18\%$, with macro-averaged precision, sensitivity, specificity, and F1-score of $92.04\%$, $83.66\%$, $97.64\%$, and $87.38\%$, respectively, as reported in Table~\ref{tab:ablation-study}. The high specificity indicates that the model effectively suppresses false positives across all arrhythmia classes, while the F1-score of $87.38\%$ reflects consistent performance despite the inherent class imbalance in the MIT-BIH Arrhythmia dataset.

\begin{figure}[!t]
    \centering
    \subfloat[ROC curve]{
        \includegraphics[width=0.75\columnwidth]{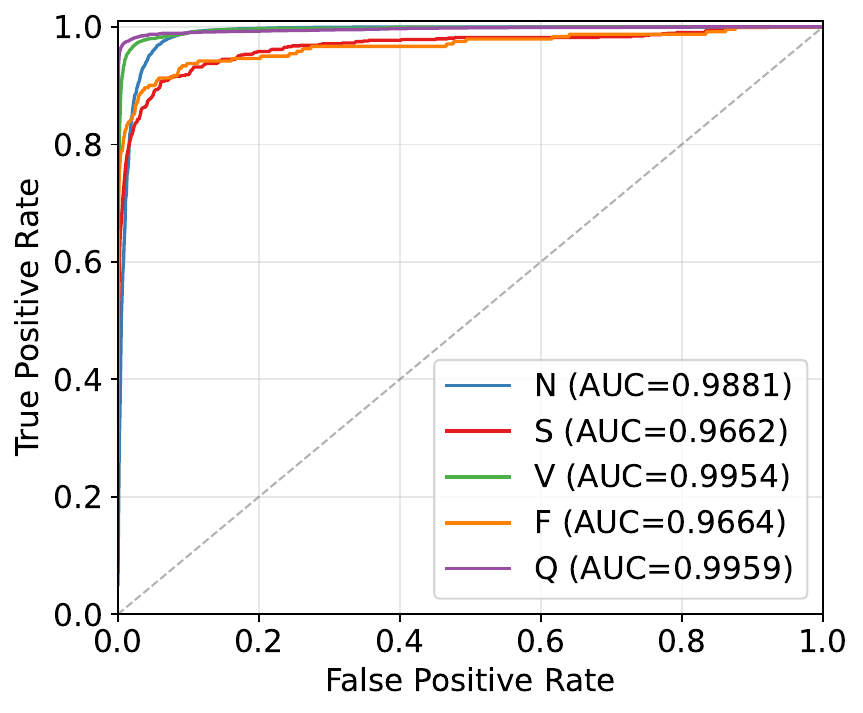}
        \label{fig:roc}
    }
    \vspace{0.5mm}
    \subfloat[PR curve]{
        \includegraphics[width=0.75\columnwidth]{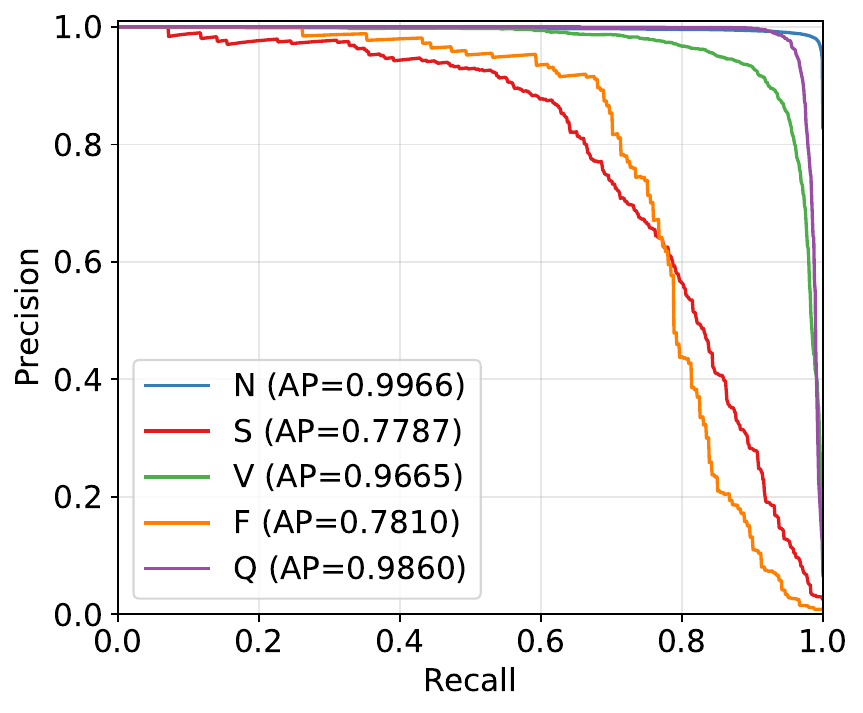}
        \label{fig:pr}
    }
    \caption{Per-class ROC and PR curves.}
    \label{fig:curves}
\end{figure}
To analyze the performance of LDC-ECG beyond hard classification metrics, the ROC and PR curves are presented. As illustrated in Fig.~\ref{fig:roc}, LDC-ECG achieves AUC scores exceeding $0.96$ across all five arrhythmia classes, reflecting strong discriminative capacity. Notably, the V and Q classes attain the highest AUC scores of $0.9954$ and $0.9959$, likely attributable to their greater representation in the training data. The S and F classes achieve AUC scores of $0.9662$ and $0.9664$, demonstrating good separability despite their limited sample sizes. The PR curves in Fig.~\ref{fig:pr} further reveal the per-class detection behavior. The N, V, and Q classes achieve high AP scores of $0.9966$, $0.9665$, and $0.9860$, respectively, maintaining high precision across the full recall range. The S and F classes yield lower AP scores of $0.7787$ and $0.7810$, consistent with the lower sensitivity and their extremely limited representation in the dataset. Nevertheless, these two classes still demonstrate a reasonable precision-recall trade-off, and decision threshold calibration can be applied to adapt to specific clinical requirements.

\begin{table}[t]
\caption{Comparison with lightweight SOTA arrhythmia classifiers on the MIT-BIH Arrhythmia Database.}
\label{tab:tinyml_comparison}
\centering
\footnotesize
\setlength{\tabcolsep}{3pt}
\renewcommand{\arraystretch}{1.2}
\begin{tabular}{lccccccccc}
\hline
\textbf{Work} & \textbf{Year} & \textbf{Weights} &
 \textbf{Acc} &
\textbf{Model Size} & \textbf{Memory} \\
& & &
(\%) & (kB) & (kB) \\
\hline
Xu et al. \cite{9131693} & 2020 & --- & 95.90 & --- & --- \\
Scrugli et al. \cite{9656169} & 2022 & 8-bit & 96.98 & --- & --- \\
Wang et al. \cite{9851002} & 2022 & 1-bit & 95.67 & 10.62 & 117.70 \\
Ahmed et al. \cite{math11030562} & 2023 & 32-bit & 99 & --- & 2200 \\
Pu et al. \cite{10193930} & 2023 &  1-bit & 96.90 & --- & 3.76 \\
Iqbal et al. \cite{10964374} & 2025 & 8-bit & 96.41 & --- & 45$\sim$180 \\
Busia et al. \cite{10531812} & 2025 & 8-bit & 98.97 & 6.64 & 49 \\
Nikandish et al. \cite{11435619} & 2026 & 8-bit  & 94.60 & 83.97 & 139 \\
\textbf{ECG-LDC (our)} & 2026 & 1-bit & 97.18 & 3.38 & 3.86 \\
\hline
\end{tabular}
\end{table}
We also benchmark ECG-LDC against state-of-the-art TinyML solutions evaluated on the MIT-BIH Arrhythmia Database to demonstrate its competitiveness as a lightweight classifier
Table~\ref{tab:tinyml_comparison} shows that ECG-LDC achieves the highest classification accuracy of $97.18\%$ among all 1-bit models and outperforms most 8-bit solutions as well, while maintaining a memory footprint of only $3.86~\text{kB}$, on par with the most memory-efficient competing method, Pu et al.~\cite{10193930} ($3.76~\text{kB}$). Compared to Busia et al.~\cite{10531812}, which achieves $98.97\%$ with 8-bit weights, ECG-LDC trades only $1.79\%$ in accuracy for a nearly $2\times$ reduction in model size and a $12.7\times$ reduction in total memory footprint, making it particularly well-suited for resource-constrained embedded deployment. Notably, Iqbal et al.~\cite{10964374} incur memory costs ranging from $45$ to $180~\text{kB}$ despite using 8-bit quantization, whereas ECG-LDC demands at least $11\times$ less memory. These results demonstrate that ECG-LDC achieves a compelling accuracy-efficiency trade-off, delivering near state-of-the-art accuracy with a fraction of the memory and storage overhead, which is critical for real-time arrhythmia detection on TinyML hardware.

\subsection{Ablation Study on Feature Selection and Codebook Design}
\begin{table}[t]
\caption{Macro-metric comparison for different feature and encoder configurations.}
\label{tab:ablation-study}
\centering
\footnotesize
\setlength{\tabcolsep}{3pt}
\renewcommand{\arraystretch}{1.2}
\begin{tabular}{lcccccc}
\hline
\textbf{Features} & \textbf{Codebook} & \textbf{Acc} & \textbf{Prec} & \textbf{Sens} & \textbf{Spec} & \textbf{F1} \\
 & & (\%) & (\%) & (\%) & (\%) & (\%) \\
\hline
RR-interval & --- & 86.81 & 60.85 & 41.19 & 86.96 & 45.29 \\
Morphological & --- & 95.52 & 90.02 & 73.13 & 95.70 & 79.68 \\
Combined & Shared & 96.66 & 91.18 & 79.19 & 97.28 & 83.94 \\
Combined & Separate & 97.18 & 92.04 & 83.66 & 97.64 & 87.38 \\
\hline
\end{tabular}
\end{table}
As this work proposes a dual-encoder extension of LDC for ECG arrhythmia classification -- incorporating separate encoders for morphological and RR-interval features -- this section presents an ablation study to systematically evaluate the contribution of each design choice. The four configurations examined are: (1) RR-interval-only, (2) morphological-feature-only, (3) combining both feature types with shared feature and value codebooks, and (4) the proposed ECG-LDC architecture with dedicated codebooks per feature component. The results in Table~\ref{tab:ablation-study} show three key observations. First, morphological features alone substantially outperform RR-interval features across all metrics, with F1-scores of $79.68\%$ versus $45.29\%$, highlighting that morphological patterns carry far more discriminative information. Second, combining both feature types improves performance, confirming that LDC can still extract complementary information from RR-intervals despite their lower individual performance. Third, the proposed architecture with distinct codebooks per feature component achieves the best performance across all metrics, particularly improving sensitivity by $4.47\%$ and F1-score by $3.44\%$. This suggests that decoupling the encoding space allows each encoder to build representations better aligned with its own feature type, reducing representation contention imposed by a shared one.

\subsection{Hardware Efficiency}
\begin{figure}[!t]
    \centering
    \includegraphics[width=0.85\columnwidth]{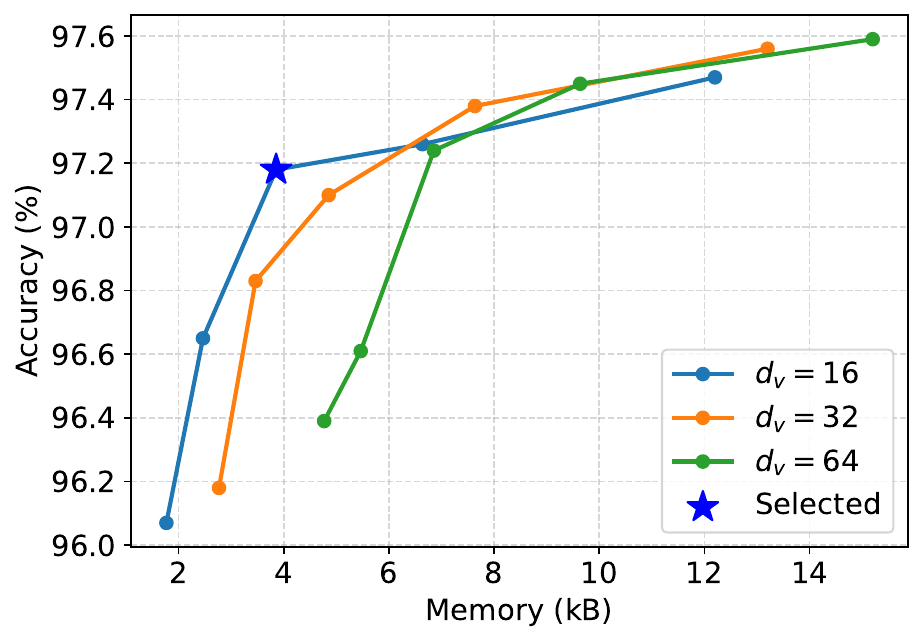}
    \caption{Accuracy--memory tradeoff of ECG-LDC}
    \label{fig:tradeoff}
\end{figure}
To understand the trade-off between classification performance and hardware efficiency, and to identify the optimal configuration, we conduct a parametric sweep across the hypervector design space. To limit the search space, we enforce $d_{v,r} = d_{v,m} = d_v$ and $d_{f,r} = d_{f,m} = d_f$, reducing the number of independent variables. In this experiment, memory usage is adopted as the primary efficiency metric. Although it does not fully capture all hardware aspects such as power consumption and latency, it directly correlates with these factors, particularly in FPGA implementations where larger memory requirements typically demand more complex routing and logic overhead. The value hypervector dimension $d_v$ is swept over $\{16, 32, 64\}$, while the feature hypervector dimension $d_f$ is swept over $\{64, 128, 256, 512, 1024\}$. The resulting trade-off curves are presented in Fig.~\ref{fig:tradeoff}. As illustrated, increasing $d_f$ consistently improves classification accuracy across all values of $d_v$, though the gain becomes marginal at larger $d_f$ values. Furthermore, smaller $d_v$ values tend to scale more efficiently in accuracy with increasing memory, while larger $d_v$ values require substantially more memory to reach comparable performance. This suggests a more favorable trade-off at lower $d_v$ settings. Notably, the selected configuration (marked by the star) achieves $97.18\%$ accuracy at only $3.86~\text{kB}$ of memory, making it a practical choice for resource-constrained IoT deployment. This result justifies the configuration in Table~\ref{tab:params}.

\begin{table}[t]
\caption{Resource utilization and performance summary.}
\label{tab:resource_breakdown}
\centering
\footnotesize
\setlength{\tabcolsep}{4pt}
\renewcommand{\arraystretch}{1.2}
\begin{adjustbox}{width=\columnwidth}
\begin{tabular}{lccc}
\hline
\textbf{Metric} & \textbf{AXI Subsystem} & \textbf{ECG-LDC} & \textbf{Full System} \\
\hline
LUT &  2589 & 5652 & 8255 \\
FF &  3473 & 4790 & 8296 \\
BRAM &  2 & 4 & 6 \\
DSP &  0 & 0 & 0 \\
Static Power (mW)  & --- & 105 & 138 \\
Dynamic Power (mW) & 10 & 34 & 1571 \\
Latency ($\mu$s) & --- & 23.9 & 4195.22 \\
Throughput (beat/s) & --- & --- & $24.43 \times 10^3$ \\
\hline
\end{tabular}
\end{adjustbox}
\end{table}

\begin{table*}[b]
\caption{Comparison with existing FPGA-based ECG arrhythmia classification approaches.}
\label{tab:fpga_comparison}
\centering
\footnotesize
\setlength{\tabcolsep}{4pt}
\renewcommand{\arraystretch}{1.2}
\begin{threeparttable}
\begin{tabular}{lcccccccccccc}
\hline
\textbf{Work} & \textbf{Year} & \textbf{Classes} & \textbf{Platform} & \textbf{System} &
\textbf{Acc} & \textbf{Freq} & \textbf{LUT} & \textbf{FF} &
\textbf{BRAM} & \textbf{DSP} & \textbf{Latency} & \textbf{Power} \\
& & & & &
(\%) & (MHz) & & & & & (ms) & (W) \\
\hline
Zairi et al. \cite{Zairi2020} & 2020 & 2 & Artix-7 & PL & 95.00 & 98.21\textsuperscript{$\dagger$} & 7598 & 2763 & --- & 214 & --- & ---\\
Jaramillo et al. \cite{JaramilloRueda2020} & 2020 & 2 & Artix-7 & PL & 94.00 & 25.5\textsuperscript{$\dagger$} & 11856 & 1664 & --- & 28 & 1.358 & --- \\
Chen et al. \cite{9629700} & 2021 & 2 & Artix-7 & PS+PL & 93.60 & 100 & 5981 & 3128 & --- & 53 & 0.257 & --- \\
Rawal et al. \cite{RAWAL2023104865} & 2023 & 4 & Zynq ZCU106 & PS+PL & 86.37 & 1.25 & 184450 & 57855 & 251 & 0 & --- & 0.63  \\ 
Liu et al. \cite{Liu2025} & 2025 & 5 & Zynq 7Z020 & PS+PL & 96.55 & 50 & 13726 & 18379 & 5.5  & 9 & 63 & 1.78 \\
Greco et al. \cite{GRECO2025101705} & 2025 & 5 & Zybo Z7-20 & PS+PL & 91.90 & 50 & 20198 & 12933 & 50.5 & 98 & ???\textsuperscript{$\ddagger$} & 1.80 \\
\textbf{ECG-LDC (our)} & 2026 & 5 & Pynq-Z2 & PS+PL & 97.18 & 50 & 8255 & 8296 & 6 & 0 & 4.1952 & 1.71 \\
\hline
\end{tabular}
\begin{tablenotes}
\footnotesize
\item[$\dagger$] Reported as the maximum achievable clock frequency rather than the operating frequency.

\item[$\ddagger$] Greco et al. report a single-sample latency of 0.000259~ms, equivalent to less than 13 clock cycles at 50~MHz. This appears unusually low given the computational complexity of the reported architecture. As we were unable to reproduce this figure and respectfully note our reservations regarding its measurement, it is omitted from direct comparison.
\end{tablenotes}
\end{threeparttable}
\end{table*}

Table~\ref{tab:resource_breakdown} summarizes the resource utilization and real-time performance of ECG-LDC (main configuration) on the Pynq-Z2 platform. The ECG-LDC core demonstrates a remarkably efficient hardware implementation, consuming only $5652$ LUTs, $4790$ FFs, and $4$ BRAMs, with zero DSP blocks. This indicates that the design avoids any multiply-accumulate-heavy operations, reflecting the lightweight, addition-based nature of LDC models. 
Integrating the AXI communication subsystem introduces additional resource overhead, bringing the full system to $8255$ LUTs, $8296$ FFs, and $6$ BRAMs, while static and dynamic power remain at $138~\text{mW}$ and $1571~\text{mW}$ respectively, well within the budget of an embedded medical platform. 
The end-to-end latency to process one complete input waveform is $4195.22~\mu\text{s}$, approximately $175\times$ the core latency of $23.9~\mu\text{s}$; the gap is largely attributed to AXI interface and handshaking overhead. However, AXI Stream enables continuous pipelined data transfer, effectively hiding this overhead by allowing new waveforms to be fed into the accelerator before the previous result is returned, resulting in a throughput of $24.43 \times 10^3$~beat/s. To put this in clinical context, standard ECG acquisition systems, such as the widely used MIT-BIH Arrhythmia Database recordings, operate at a sampling rate of around $360~\text{sample/s}$. The accelerator comfortably outpaces the incoming data rate by roughly two orders of magnitude, leaving ample headroom for multi-lead monitoring, system overhead, or duty-cycled low-power operation.

Table~\ref{tab:fpga_comparison} benchmarks ECG-LDC against existing FPGA-based ECG  arrhythmia classifiers to demonstrate its competitiveness in terms of resource utilization, latency, and power efficiency. It is worth noting that although the accelerator is operated at $50~\text{MHz}$, this is not a hardware limitation but rather a deliberate design choice. The architecture can comfortably sustain frequencies up to $100~\text{MHz}$; however, the chosen frequency is sufficient to meet real-time processing requirements, while operating at a lower frequency directly reduces dynamic power consumption. ECG-LDC achieves the highest classification accuracy across all listed approaches. In terms of resource utilization, ECG-LDC is the least demanding among all five-class implementations, using over $1.6\times$ fewer LUTs and more than $2.2\times$ fewer FFs than the closest five-class competitor, Liu et al.~\cite{Liu2025}, while being the only five-class solution that requires zero DSP blocks. While earlier two-class designs such as Zairi et al.~\cite{Zairi2020} and Chen et al.~\cite{9629700} report lower LUT and FF counts, this is naturally expected given that their classification task is generally less complex. Regarding latency, the end-to-end latency of $4.1952~\text{ms}$ is higher than some competing works. Notably, Jaramillo et al.~\cite{JaramilloRueda2020} report lower latency figures; however, these measurements reflect PL-only implementations and therefore represent a different reporting scope, rather than a strictly lower full system latency. Additionally, as discussed previously, the AXI Stream interface yields a throughput of $24.43 \times 10^3~\text{beat/s}$, far exceeding the real-time processing requirements of ECG arrhythmia classification. Finally, ECG-LDC achieves the lowest power consumption among all full PS+PL five-class systems at $1.71~\text{W}$. 

\section{Conclusion}
In conclusion, this work presents ECG-LDC, a novel low-dimensional computing architecture for real-time arrhythmia classification on resource-constrained wearable devices. ECG-LDC employs a dual encoder module that independently processes morphological and RR-interval features of the ECG signal, enabling effective capture of both intra-beat waveform characteristics and inter-beat temporal dynamics. Through comprehensive experiments, ECG-LDC demonstrates effectiveness in arrhythmia classification while achieving high efficiency in terms of memory usage, inference latency, power consumption, and resource utilization. The results further establish its competitiveness against existing TinyML solutions, affirming its practicality for edge deployment.

Despite the aforementioned contributions, this work acknowledges several limitations. First, the presented system utilizes the built-in AXI Stream interface, which is convenient for rapid prototyping; however, the general-purpose AXI modules may introduce redundant logic, leading to unnecessary latency and resource overhead. Second, as LDC can be regarded as a form of hyperdimensional computing, a paradigm known for its compositional and structurally interpretable representations, this aspect of ECG-LDC has not been sufficiently addressed in this work. Future investigation of the learned representation space may help explain how discriminative arrhythmia-related patterns are structured within the low-dimensional hypervectors.

The primary objective of this work is to introduce the ECG-LDC architecture and evaluate its practicality in a real-time wearable arrhythmia classification task. Within the scope of this work, an FPGA is used as the primary platform for assessing hardware efficiency. Future work may explore deployment to microcontroller units (MCUs) or even custom application-specific integrated circuits (ASICs) to achieve greater computational and energy efficiency. Furthermore, this study is restricted to 5-class AAMI arrhythmia classification using single-lead ECG signal; future efforts may extend this to a wider range of settings, including alternative class separations, multi-lead data, and other clinically relevant tasks. Finally, this work employs a coarse-grained parametric sweep with constrained and simplified configurations to identify a balanced accuracy-efficiency operating point. Future work could leverage more advanced design space exploration strategies, such as multi-objective optimization or neural architecture search, to enable finer-grained and broader exploration of the design space.


\balance


\bibliographystyle{IEEEtran}
\bibliography{references}

\end{document}